\documentclass{article}

\usepackage{arxiv}
\usepackage[T1]{fontenc}
\usepackage[colorlinks=true, citecolor=blue]{hyperref}
\usepackage[square, numbers, sort&compress]{natbib}
\usepackage{
	bm,
	dcolumn,
	graphicx,
	float,
	subcaption,
	physics,
	tikz,
	etoolbox,
	url,
	booktabs,
	amsfonts,
	nicefrac,
	microtype,
	cleveref,
	doi,
  	epigraph,
}

\usepackage{algorithm}
\usepackage{algpseudocode}

\usepackage[base]{babel}

\graphicspath{{figures/}}

\DeclareMathOperator{\diag}{diag}

\newsavebox{\orcidbox}
\sbox{\orcidbox}{\includegraphics[width=1em]{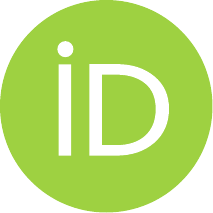}}
\newcommand{\orcid}[1]{%
	\href{https://orcid.org/#1}%
	{\usebox{\orcidbox}}\,}

\newsavebox{\scholarbox}
\sbox{\scholarbox}{\includegraphics[width=1em]{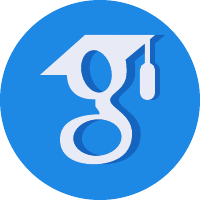}}
\newcommand{\scholar}[1]{%
	\href{https://scholar.google.com/citations?user=#1}%
	{\usebox{\scholarbox}}\,}

\title{Inverse scattering transform via affine map: applications to high-speed\\ nonlinear optical communications}
\renewcommand{\shorttitle}{Inverse Scattering Transform via Affine Map}

\date{%
	\small
	Published online in \textit{Physica D: Nonlinear Phenomena}\\
	\href{https://doi.org/10.1016/j.physd.2026.135344}
	{doi: 10.1016/j.physd.2026.135344}
}

\newif\ifuniqueAffiliation
\uniqueAffiliationtrue

\ifuniqueAffiliation 
\author{
	\scholar{B7RqVKYAAAAJ}\orcid{0009-0006-7242-7291}%
	Ilia Kuk\\
	Department of Mathematics\\
	The University of Arizona\\
	Tucson, AZ 85721, USA\\
	\texttt{ilyakuk@arizona.edu}\\
	\And
	Ildar R. Gabitov\\
	Department of Mathematics\\
	The University of Arizona\\
	Tucson, AZ 85721, USA\\
	\texttt{gabitov@arizona.edu}\\
}
\else
\usepackage{authblk}

\author[1]{%
	\scholar{B7RqVKYAAAAJ}\orcid{0009-0006-7242-7291}%
	Ilia Kuk\thanks{ilyakuk@arizona.edu}}
\author[1]{%
	Ildar R. Gabitov}

\affil[1]{Department of Mathematics, %
	The University of Arizona, %
	Tucson, AZ 85721, USA}
\fi

\hypersetup{
pdftitle={affine-map-preprint},
pdfauthor={Ilia Kuk},
}

\begin{document}
\maketitle

\begin{abstract}
	This work presents an affine map approximation for solving the inverse scattering problem related to the nonlinear Schr\"odinger model of signal propagation in high-speed coherent optical communications. Numerical simulations indicate that accurate recovery of the transmitted bit sequence can be achieved using only the continuous part of the Lax spectrum at the fiber output, thereby allowing the discrete (soliton) spectrum to be disregarded. We observed that the numerically evaluated rank of the resulting affine map matrix equals the number of bits per transmitted sequence, and we utilize this to derive a reduced order affine map.
\end{abstract}


\section{Introduction}

Vladimir E. Zakharov was a foundational figure whose seminal contributions profoundly influenced the development of modern nonlinear science across a broad spectrum of areas, ranging from cosmology and nonlinear waves to turbulence, nonlinear optics, and differential geometry. A notable example is his  contribution to the Theory of Integrable Systems. The impact of this theory  was aptly summarized by Fields Medalist Sergei Novikov: ``A large part of the most important discoveries in mathematics and in the mathematical methods of physics was made in the process of developing the theory of integrable models''~\cite{novikov1992integrability}. In particular, Vladimir Zakharov played a significant role in the development of the inverse scattering transform (IST)~\cite{zakharov1979integration, novikov1984theory, ablowitz1981solitons, newell1985solitons}, a powerful method within the mathematical framework of integrable systems, designed to solve a specific class of nonlinear partial differential equations (PDEs) known as integrable equations. Integrating nonlinear equations via the IST is a well-documented procedure in the literature. It reformulates the original problem as the evolution of scattering data for an associated spectral problem, then reconstructs the potential from those data via the inverse scattering step. The reconstructed potential satisfies the original nonlinear equation.

Notably, as demonstrated in~\cite{ablowitz1974inverse}, this scheme reduces to the Fourier transform in the linear limit. Thus IST can be considered as a nonlinear generalization of Fourier transform. The main tools for solving the inverse scattering problem are the Gelfand-Levitan-Marchenko (GLM) integral  equations and the Riemann-Hilbert (RH) problem approach. Notably, both the GLM and RH formulations are linear, despite being used to solve nonlinear problems.

In this paper, we explore the use of affine map to construct a surrogate for the inverse scattering problem related to the nonlinear Schr\"odinger equation (NLSE) as a representative example in the context of fiber-optic communication systems. This choice is motivated by the fact that integration of the NLSE marked  first major achievement by Zakharov and Shabat in the development of IST, and that the NLSE remains one of the most important integrable models with broad applicability across various physical domains.

The widespread adoption of the NLSE as a mathematical model for describing the propagation of optical pulses in communication systems from the late 1980s to the late 1990s was historically driven by three complementary factors: the development of inverse scattering transform and its application to the integration of the NLSE~\cite{zakharov1973interaction}; the demonstration that  NLSE effectively models pulse propagation in fiber-optic systems~\cite{hasegawa1973transmission, hasegawa1990guiding}; and the experimental validation of these theoretical predictions in seminal works such as~\cite{mollenauer1980experimental, mollenauer1996demonstration}. These foundational contributions induced intensive global research in the field.

The primary motivation driving research into soliton-based transmission was the remarkable stability of the soliton solutions obtained in~\cite{zakharov1973interaction}, which suggested their suitability as robust bit carriers in optical fiber systems. The topic of soliton communications has led to the solution of many experimental, theoretical and practical challenges. In the mid-90s, the possibilities of soliton communication were demonstrated at the international exhibitions CEBIT 97 in Hannover and EXPO 98 in Lisbon (on the Madrid-Lisbon line) within the framework of the European project UPGRADE, where in the conditions of outdated fiber links, the productivity of conventional technologies at that time was exceeded by 4 times.

During the evolution of this field, a more effective strategy emerged under the name ``dispersion management,'' in which the standard nonlinear Schr\"odinger model is modified so that the dispersion coefficient alternates periodically in sign along the fiber~\cite{knox199510, gabitov1996averaged, nakazawa1995construction, gabitov1996breathing, mollenauer2006solitons}. This method was industrially implemented as a fiber optic line based on the same principle, deployed in Australia~\cite{pratt20035}. Notably, Zakharov and Manakov~\cite{zakharov1999propagation} showed that the equations governing dispersion-managed pulse dynamics as well as the standard NLSE are integrable. To our best knowledge this challenge is still not addressed properly.

In the early 2000s, advances in hardware and signal processing techniques gave rise to coherent communication as a new technology for fiber-optic data transmission. Its main advantage was an order-of-magnitude improvement in spectral efficiency, enabling a larger symbol alphabet within each bit slot. The implementation of this approach required a change in the modulation format, in which information was encoded by the phase difference of adjacent pulses. The disadvantage of such a system is in its fundamentally linear nature. With a further increase in the transmission speed of optical fiber systems and the transmission length, an increase in the negative impact of the nonlinearity of the optical fiber is inevitable. This challenge, viewed from a modern perspective, revives the issues encountered in the 1990s and underscores the need to generalize the principles of coherent communication to the nonlinear regime of optical pulse propagation.

Over the last decade, the application of the inverse scattering transform (also known as the nonlinear Fourier transform, NFT) to fiber-optic communications has developed into a broad research direction motivated by the need for nonlinearity mitigation in Kerr-limited links. The survey by Turitsyn et al.~\cite{turitsyn2017nonlinear} frames NFT-based transmission as a paradigm in which nonlinear propagation is handled in a spectral domain tailored to the integrable model, and provides a unified review of system designs, spectral efficiency considerations, and the practical bottlenecks that arise from numerical accuracy and complexity of the direct and inverse transforms. Complementing this perspective, Sedov et al.~\cite{sedov2025numerical} discuss the role of NFT/IST methods in telecommunications with a focus on efficient numerical algorithms for implementation, both for channel equalization and for time-domain signal processing tasks.

In parallel, machine learning has been explored as a data-driven mechanism to approximate or augment difficult steps in IST processing. Sedov et al.~\cite{sedov2021neural} propose a structured neural architecture to compute the continuous nonlinear spectrum for decaying profiles, emphasizing improved robustness for signals corrupted by noise. Other works use neural networks for receiver side post-processing of the nonlinear spectrum to improve detection performance and compensate implementation penalties: Kotlyar et al.~\cite{kotlyar2020combining,kotlyar2021convolutional} study neural post-processing of the nonlinear spectrum and introduce recurrent and convolutional equalizers designed to capture correlations across spectral components.

\section{Theoretical Background}

\subsection{Nonlinear Schr\"odinger equation}

An optical pulse is characterized by a duration $\tau_0$ and an electric field $u$. The main characteristics of an optical fiber include the chromatic dispersion $\beta$, the attenuation (loss) coefficient $\alpha$, and the Kerr nonlinearity coefficient $\gamma$. In standard fiber-optics notation, the Kerr effect gives an intensity-dependent refractive index $n = n_0 + n_2 I$, which leads to a nonlinear phase accumulation captured by the parameter $\gamma$ in the envelope model, where $n_0$ is the linear refractive index and $I=|u|^2$ is the electric field intensity. The characteristic scales at which the dispersion and nonlinearity effects become noticeable are $Z_D=\tau_0^2/\beta$ and $Z_{\mathrm{nl}}=(\gamma I)^{-1}$, respectively. In-line optical amplifiers are incorporated into optical links to compensate for fiber losses; the characteristic distance between amplifiers $Z_a \propto \alpha^{-1}$ is set by the attenuation $\alpha$. The evolution of optical pulses in fiber lines with $Z_{\mathrm{nl}}, Z_D \gg Z_a$, after appropriate normalization, can be described with sufficient accuracy by the nonlinear Schr\"odinger equation, under the conditions described in \cite{hasegawa1990guiding, mollenauer2006solitons}:
\begin{equation}
	i u_z + \tfrac12 u_{tt} + |u|^2u = 0,
	\label{eq:nlse}
\end{equation}
where the nominal value of the electric field $u$ is the result of path-averaging over one amplifier span of length $Z_a$. Here distributed loss and lumped amplification are averaged over one span, yielding an effective conservative evolution for the normalized envelope; this path-averaged model neglects span-to-span power variations and does not capture effects such as amplified spontaneous emission noise. Equation~\eqref{eq:nlse} is written in the focusing normalization (corresponding to anomalous dispersion in dimensional variables); the dispersion sign distinguishes anomalous and normal dispersion regimes.

For the implementation of modulation formats in fiber-optic communication, time is conventionally divided into bit slots of duration $T$. Until the early 2000s, amplitude modulation formats were predominantly used. In this case, the slot containing the optical pulse corresponded to the logical ``1'', the empty slot corresponded to the logical ``0''. In the return-to-zero (RZ) format, the optical field intensity rises from zero to its peak and then returns to zero within each bit slot. In contrast, the non-return-to-zero (NRZ) format holds the field at a constant level for the duration of consecutive logical ``1'' slots, without returning to zero between bits.

Currently, more advanced modulation formats have become widespread, in which information is encoded based on the phase difference between adjacent pulses. In the simplest differential phase-shift keying (DPSK) each pulse is assigned a phase of either $0$ or $\pi$. A generalization of this scheme is quadrature phase-shift keying (QPSK), where each symbol represents two bits of information. In QPSK, the phase can assume one of four discrete values (e.g., $0$, $\pi/2$, $\pi$, or $3\pi/2$), thereby doubling the spectral efficiency relative to binary phase modulation.

These modulation schemes can be conveniently illustrated on the complex plane using constellations represented by points on a circle. The radius of the circle corresponds to the characteristic amplitude of the optical pulse. In DPSK, pulses map to two points, $\pm 1$, on the unit circle, assuming normalized amplitude. In QPSK, pulses correspond to four points on the unit circle given by $\exp(i \pi k/2),\quad k = 0, 1, 2, 3$. Nowadays, the telecommunications industry employs more sophisticated techniques to utilize the bandwidth of optical fibers. In addition to increasing the number of phase states (i.e., pairs of pulses with opposite phases), modulation schemes also exploit the flexibility in selecting the amplitude of the pulses. As a result, the signal constellation can be visualized as a regular grid of points within a rectangular region in the complex plane~\cite{kikuchi2016fundamentals}. A detailed analysis of such systems lies beyond the scope of this discussion.

\begin{figure}
	\centering
	\begin{subfigure}{0.35\linewidth}
		\centering
			\includegraphics[width=\textwidth]{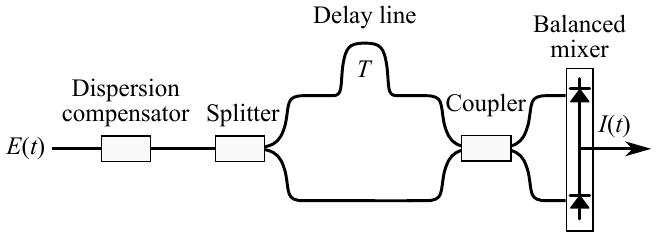}
		\caption{Direct detection}
		\label{fig:detection-direct}
	\end{subfigure}
	\begin{subfigure}{0.35\linewidth}
		\centering
		\includegraphics[width=\textwidth]{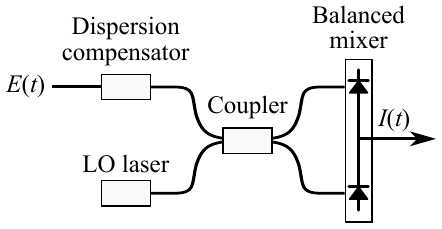}
		\caption{Coherent detection}
		\label{fig:detection-coherent}
	\end{subfigure}
	\caption{Detection schemes}
\end{figure}

The  dispersion broadening of the pulses accumulated during the passage of the entire line is subject to digital compensation at the receiving terminal. There are two main methods used to decode the signal: ``direct'' or ``coherent'' decoding. In the case of direct decoding, the pulse stream is divided into two identical copies, after which one of the copies is delayed by a bit slot and added to the second copy. If the phases of the summed pulses coincide, the result of constructive interference is an optical pulse in the corresponding slot. Otherwise, the slot will be empty as a result of destructive interference. Thus, the differential phase modulation is converted into amplitude modulation which is then converted into a bit sequence after registration by the photo-diode (see Fig.~\ref{fig:detection-direct}). In coherent decoding, the pulse stream after dispersion compensation is mixed with a continuous beam of the reference laser, the phase of which is locked with the transmitting laser at the beginning of the communication line. After mixing, the differential phase modulation is transformed into amplitude modulation on a balanced mixer, as shown in Fig.~\ref{fig:detection-coherent}.

Coherent communication technique is currently the main technology for high-speed communication~\cite{kikuchi2016fundamentals}. However, its applicability is limited by the requirement of linear propagation of optical signal in optical fiber. One of the most pronounced manifestations of nonlinearity is nonlinear phase distortion, which leads to errors in information transmission. Extension of coherent transmission principles for systems operating in nonlinear mode is a serious challenge to modern high-speed communication. In this work, we consider one of the possible ways to address this problem based on the use of the integrability property of the basic model and the numerical solution of the inverse scattering problem for a special class of potentials corresponding to the RZ modulation format using affine map operator.

\subsection{Fundamentals of the Inverse Scattering Transform}

The mathematical foundation of IST was laid by Peter Lax in 1968~\cite{lax1968integrals}, who introduced a formalism to associate a nonlinear PDE with a pair of linear operators $L$ and $P$, today known as a ``Lax pair'', which satisfy
\begin{equation*}
	L_t + L P - P L = 0.
\end{equation*}

The NLSE~\eqref{eq:nlse} admits a Lax pair representation~\cite{shabat1972exact} in the following form
\begin{equation}
	\begin{aligned}
		\psi_t &= i\bigl(\lambda \sigma_z + U\bigr) \psi, \\
		\psi_z &= i\bigl(-\lambda^2 \sigma_z - \lambda U + V\bigr) \psi,
	\end{aligned}
	\label{eq:lax-pair-eq}
\end{equation}
where $\psi$ is a $2\times2$ matrix-valued function, $\lambda = \xi + i \zeta$ is the complex spectral parameter, and $\sigma_z = \diag(1, -1)$ is the Pauli matrix. 

Introducing matrices $U$ and $V$ which depend on the electric field $u(z,t)$ as
\begin{equation*}
	U =
	\begin{bmatrix}
		0 & u \\
		u^* & 0
	\end{bmatrix}, 
	\quad
	V = \tfrac12
	\begin{bmatrix}
		|u|^2 & i u_t \\
		- i u_t^* & -|u|^2
	\end{bmatrix}.
\end{equation*}
The compatibility condition $\psi_{tz} = \psi_{zt}$ leads directly to the Zakharov-Shabat equation
\begin{equation*}
	U_z - V_t + [ U, V ] = 0,
\end{equation*}
which is equivalent to the NLSE.

To simulate signal propagation in optical fibers, we consider the direct scattering problem associated with the first operator of the Lax pair. For solutions with given asymptotics, known as Jost functions, we have
\begin{equation*}
	\lim_{t \to -\infty} \psi =
	\begin{bmatrix}
		e^{-i\lambda t} \\
		0
	\end{bmatrix}, 
	\quad
	\lim_{t \to +\infty} \psi =
	\begin{bmatrix}
		a(z,\lambda) e^{-i\lambda t} \\
		b(z,\lambda) e^{+i\lambda t}
	\end{bmatrix}.
\end{equation*}

The scattering data associated with the Zakharov-Shabat problem in general separates into a continuous and a discrete part. For $\lambda$ on the real axis ($\lambda=\xi\in\mathbb{R}$), one obtains the continuous spectrum, which is commonly represented by the reflection coefficient $r(z,\xi)$. In addition, the discrete spectrum consists of isolated eigenvalues $\{\lambda_k\}$ in the upper half-plane ($\Im\lambda_k>0$) at which $a(z,\lambda_k)=0$; these correspond to soliton components of the field and are accompanied by associated norming constants. Thus, the full nonlinear spectrum is given by the reflection coefficient on the real axis together with the discrete eigenvalues and their norming data.

The reflection coefficient (continuous part of scattering data) is defined as
\begin{equation}
	r(z,\lambda) = \frac{b(z,\lambda)}{a(z,\lambda)}.
	\label{eq:ref-coef}
\end{equation}

The evolution of this  coefficient along space $z$  obeys the linear Schr\"odinger equation. The linear nature of the dynamics of the reflection coefficient allows to suppress nonlinear distortions by digital compensation of the accumulated dispersion of not the physical field but the scattering coefficient, see for example~\cite{prilepsky2014nonlinear}.

\subsection{Brief overview of nonlinear effects in optical fibers}

In addition to self-phase modulation, nonlinearity of optical fiber entails many additional effects leading to an increase in bit error rates in communication systems. These include inter- and intra-channel interactions. The study of inter-channel interactions is beyond the scope of this article. Overlapping of optical pulses in a bit sequence under the influence of dispersion in the presence of nonlinearity, in addition to phase distortions,  leads to emergence of a four-wave interaction. This interaction results in redistribution of energy along the bit sequence, amplitude jitter and generation of transmission errors~\cite{mamyshev1999pulse}.

In cases where the Lax spectrum of optical pulses contains both discrete (soliton sector of the spectrum) and continuous components, the propagation of pulses in the optical fiber is accompanied by shedding of continuous radiation. This leads to a time jitter of bit carriers, which also a  source  to errors.~\cite{chertkov2003shedding, kuznetsov1995nonlinear} This effect is in some sense a generalization of the Gordon-Haus jitter phenomenon, well known and widely discussed in the past in connection with soliton telecommunications.~\cite{gordon1986random, elgin1993stochastic}

The combination of these subtle and non-trivial nonlinear effects along with the instability of numerical algorithms for solving the direct and inverse problems for the case of potentials with a complex structure is a challenge for the implementation of the procedure described above. Discretization of the optical field at the receiving terminal  introduces numerical noise into the scattering data, which in turn demands a separate stability analysis.

The presence of a discrete spectrum in the potential under study leads to the appearance of growing exponentials, which is also a potential source of instability of numerical algorithms.~\cite{novikov1984theory} On the other hand, a continuous spectrum is a superposition of plane waves that do not contain growing exponentials. Its evolution is accompanied by a change in phase relationships. This creates prerequisites for using an incomplete basis - a continuous spectrum for recovering the RZ-type potential.

\section{Affine approximation of the inverse scattering transform}

\subsection{Integrability-based signal processing workflow}

\begin{figure}
	\centering
	\includegraphics[width=0.5\linewidth]{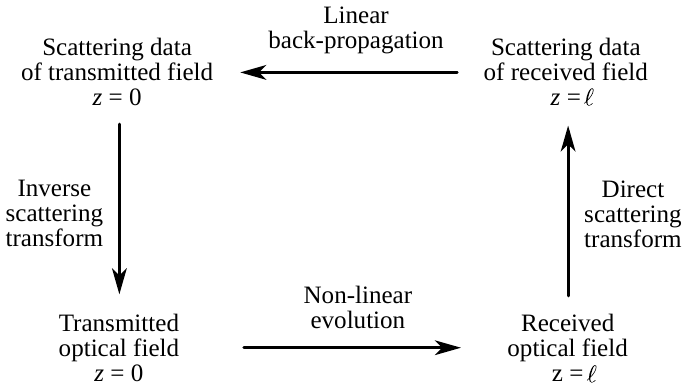}
	\caption{Integrability-based signal processing workflow scheme}
	\label{fig:idea}
\end{figure}

Figure~\ref{fig:idea}
is a schematic representation of the corresponding  signal processing workflow. At the transmitter ($z = 0$) an initial optical field is transmitted into the fiber. As it propagates through the fiber, the field evolves according to the NLSE, impacted by nonlinear distortions. At the receiver ($z = \ell$), the received optical field is processed using the Direct Scattering Transform (DST), giving us the scattering data corresponding to the received signal. Since the scattering data evolves in a simple linear way, it is then numerically backpropagated. Finally, the Inverse Scattering Transform is applied to the backpropagated scattering data, reconstructing an approximation of the original transmitted optical field. This approach enables a signal recovery by harnessing the integrability of the NLSE and the linear evolution of the associated scattering data.

\subsection{Solution map operator}

Recovering the initial potential from scattering data is challenging both analytically and numerically. Classical methods, such as the Gelfand-Levitan-Marchenko (GLM) equations and the Riemann-Hilbert (RH) problem, can involve high computational cost and complexity~\cite{trogdon2013numerical}. Here we seek for a computationally efficient and simple approach to approximate the inverse scattering transform solution operator
\begin{equation*}
	G:\mathcal{X}\to\mathcal{Y},\quad y = G(x),
\end{equation*}
where each element $x \in \mathcal{X}$ is a scattering data function $x(\xi) = r(0,\xi)$ living in the space of scattering data functions $\mathcal{X}$ and $y \in \mathcal{Y}$ is the corresponding initial potential $y(t) = u(0,t)$ living in the space of initial potential functions $\mathcal{Y}$. In other words, given the continuous spectrum profile $x(\xi)$ as a function of spectral parameter $\xi$, which is the real part of $\lambda$ defined in the Eq.~(\ref{eq:lax-pair-eq}), the operator $G$ returns the temporal profile $y(t)$ of the initial potential.  

Given a finite observations set $\{(x_j,y_j)\}_{j=1}^N\subset\mathcal{X}\times\mathcal{Y}$ with $y_j = G(x_j)$, where $N$ is the number of sample pairs, we introduce an approximation parametric operator $G_\theta:\mathcal{X}\to\mathcal{Y}$ with $\theta\in\Theta$, where $\Theta$ is a finite-dimensional parameter space, by minimizing a loss function
\begin{equation}
	\min_{\theta\in\Theta} \frac1N\sum_{j=1}^N 
	C\bigl(G_\theta(x_j), y_j\bigr),
	\label{eq:optimization}
\end{equation}
where $C:\mathcal{Y}\times\mathcal{Y}\to\mathbb{R}$ is a loss function which measures discrepancy.  In this work we choose the relative $L_2$ error,
\begin{equation}
	C(\hat y,y)=\frac{\|\hat y-y\|_{2}}{\|y\|_{2}}.
	\label{eq:loss}
\end{equation}

Although the GLM and RH frameworks are linear at the level of equations involved, in general the overall map $G$ from scattering data to initial potential is nonlinear.

\subsection{Affine map operator}

For numerical implementation, we sample each scattering data function $x_j$ and initial potential function $y_j$ at $n$ and $m$ points, respectively. This yields the row vectors
\begin{equation*}
	\mathbf{x}_j = [ x_j(\xi_1), \dots, x_j(\xi_n) ]
	\quad\text{and}\quad
	\mathbf{y}_j = [ y_j(t_1), \dots, y_j(t_m) ],
\end{equation*}
where the spectral parameter sample points $\{\xi_i\}_{i=1}^n$ cover the domain of the scattering data, and the time-domain sample points $\{t_i\}_{i=1}^m$ span the interval over which the initial potential is defined. Here $n$ and $m$ are numbers of corresponding discretization points.

We propose to approximate a solution map by a simple affine model 
\begin{equation}
	\mathbf{y}_j
	= G_\theta(\mathbf{x}_j)
	= \mathbf{x}_j A + \mathbf{b},
	\label{eq:affine-tr}
\end{equation}
where $A\in\mathbb{R}^{2n\times 2m}$ and $\mathbf{b}\in\mathbb{R}^{1\times 2m}$ constitute parameters $\theta$ and are learned from the optimization problem~(\ref{eq:optimization}). Equation~(\ref{eq:affine-tr}) is mathematically equivalent to a multivariate linear regression model with an intercept: each of the $2m$ output components is an affine function of the $2n$ input components. The affine map terminology emphasizes the viewpoint of $G_\theta$ as an operator approximating the inverse scattering solution map, rather than a statistical model for scalar prediction. In particular, we exploit that this choice enables a closed-form identification and exposes the low-rank structure of the inverse map surrogate through the Singular Value Decomposition (SVD). We use the term affine to explicitly include the bias term $\mathbf{b}$, which would be absent in a strictly linear map.

The model~\eqref{eq:affine-tr} is algebraically identical to a one-layer fully connected neural network when no activation function is applied. In that sense, the present surrogate can be viewed as the linear limit of a feedforward network architecture. The distinction is not in the functional form, but in the modeling objective and the identification procedure: we treat $G_\theta$ as a discretized operator approximation to the inverse scattering solution map and identify its parameters by a single solve~\eqref{eq:full-map}, rather than by iterative gradient-based training.
 
We collect complex-valued observations into two matrices
\begin{align*}
	X_{\mathrm{c}}
	= \begin{bmatrix}
		\mathbf{x}_1 \\
		\mathbf{x}_2 \\
		\vdots \\
		\mathbf{x}_N
	\end{bmatrix}
	\in \mathbb{C}^{N\times n}
	\quad\text{and}\quad
	Y_{\mathrm{c}}
	= \begin{bmatrix}
		\mathbf{y}_1 \\
		\mathbf{y}_2 \\
		\vdots \\
		\mathbf{y}_N
	\end{bmatrix}
	\in \mathbb{C}^{N\times m}.
\end{align*}

To enable the affine operator to capture interactions between the real and imaginary parts of functions, we split each complex-valued matrix into its real and imaginary parts and then concatenate them side-by-side forming the real-valued matrices $X \in \mathbb{R}^{N\times 2n}$ and $Y \in \mathbb{R}^{N\times 2m}$ as
\begin{equation*}
	X = [ \Re(X_{\mathrm{c}}) \Im(X_{\mathrm{c}}) ],\quad
	Y = [ \Re(Y_{\mathrm{c}}) \Im(Y_{\mathrm{c}}) ].
\end{equation*}

To unify the bias term, we define
\begin{equation*}
	X_{\mathrm{aug}}
	= [ X  \mathbf{1} ]
	\in\mathbb{R}^{N\times(2n+1)},\quad
	A_{\mathrm{aug}}
	= \begin{bmatrix}A\\\mathbf{b}\end{bmatrix}
	\in\mathbb{R}^{(2n+1)\times 2m},
\end{equation*}
so that the affine model~(\ref{eq:affine-tr}) is written as
\begin{equation*}
	Y \approx X_{\mathrm{aug}} A_{\mathrm{aug}}.
\end{equation*}

Thus, to find a discretized version of the affine operator $A_\mathrm{aug}$ we solve the least-squares problem applying the Moore-Penrose pseudo-inverse~\cite{moore1920reciprocal, penrose1955generalized} as follows
\begin{equation}
	A_{\mathrm{aug}} = X_{\mathrm{aug}}^{+} Y,
	\label{eq:full-map}
\end{equation}
where $^{+}$ denotes pseudo-inverse.

Further, we can exploit low-rank structure of the affine operator introducing a factorization
\begin{equation*}
	A_{\mathrm{aug}} \approx R A_r, 
	R \in \mathbb{R}^{(2n+1)\times r}, 
	A_r \in \mathbb{R}^{r\times 2m}, 
	r \ll 2n+1,
\end{equation*}
where $R$ and $A_r$ found via truncated singular-value decomposition
\begin{equation*}
	A_{\mathrm{aug}} = U \Sigma V^T,\quad
	R = U_r \Sigma_r,\quad
	A_r = V_r^T,
\end{equation*}
where $U_r$ and $V_r$ contain the leading $r$ left and right singular vectors of $A_{\mathrm{aug}}$ respectively and $\Sigma_r$ the corresponding singular values. The reduced model then formulated as
\begin{equation}
	Y \approx (X_{\mathrm{aug}} R) A_r
	\quad\text{or}\quad
	\mathbf{y}_j = \left(\mathbf{x}_j R\right) A_r.
	\label{eq:reduced-map}
\end{equation}

Thus, the scattering data are first projected into an $r$-dimensional subspace via the projection matrix $R$, and the reduced affine operator $A_r$ is then applied to reconstruct the initial potential.  This procedure not only lowers computational cost but also suggests a pathway toward faster numerical algorithms for the DST.

\section{Results}

We report results for two related settings, both targeting recovery of the transmitted waveform $u(0,t)$. In the idealized case, the affine map is trained and tested as a direct map from the reflection coefficient computed at the transmitter end, $x(\xi)=r(0,\xi)$, to the corresponding initial potential $y(t)=u(0,t)$. In the end-to-end setting, the waveform is first propagated by the NLSE to $z=z_{\mathrm{end}}$, the scattering data are computed at $z=z_{\mathrm{end}}$, the continuous spectrum is backpropagated to $z=0$, and the affine surrogate is applied to reconstruct an approximation of the transmitted signal.

The results in this work are reported for RZ Gaussian pulse trains with binary DPSK symbols, which provide a testbed for studying an affine surrogate of the inverse scattering map. While this modulation has comparatively low spectral efficiency, it retains key features relevant to integrability-based processing: localized initial pulses and strong nonlinear interactions under NLSE propagation.

Extending the present framework to higher spectral efficiency waveforms introduces additional structure and challenges. Nevertheless, the proposed procedure and reduced rank factorization remain applicable: the affine surrogate can be trained on input to output pairs generated from the chosen modulation format, and the singular value decay of the learned operator provides an empirical diagnostic of the intrinsic dimensionality and compression potential for particular signal class. A systematic evaluation on higher-order modulation formats and pulse shaping models requires additional study.

\subsection{Data generation and preprocessing}
\label{sec:data-gen}

We first describe the dataset used for the idealized case experiments in Sec.~\ref{sec:num-exp}, where scattering data are computed directly from the transmitted waveform at $z=0$.

We generate return-to-zero (RZ) Gaussian pulse trains with differential phase-shift keying (DPSK) encoding. For each sequence $j = 1, \dots, N$, the amplitude coefficients $a_j^k \in \{+1, -1\}$ are drawn independently with
\begin{equation*}
	\mathbb{P}\bigl(a_j^k = +1\bigr)
	=
	\mathbb{P}\bigl(a_j^k = -1\bigr)
	=
	\tfrac12,
	\qquad
	k = 1, \dots, K, 
	K = 16,
\end{equation*}
where $\mathbb{P}(\cdot)$ denotes the probability. The resulting initial potential at $z=0$ is given by
\begin{equation*}
	y_j(t)=u\left(0,t;\{a_j^k\}_{k=1}^K\right)
	=\sum_{k=1}^K a_j^k \pi^{-1/4}\exp \Bigl[-\tfrac12(t-kT)^2\Bigr],
\end{equation*}
with bit-slot width $T=10$ and normalization factor $\pi^{-1/4}$ ensuring unit pulse energy. We sample $t$ uniformly on $[-90,90]$ interval using $M_t=4096$ points.

To obtain the scattering data for each $y_j(t)$, we compute the DST numerically using the Fast Nonlinear Fourier Transform (FNFT) software package developed by Wahls and Poor, following the algorithmic framework introduced in~\cite{wahls2013introducing}. For each input waveform sampled on the uniform time grid ($M_t=4096$ points on $t\in[-90,90]$), we evaluate the Zakharov-Shabat scattering problem and extract the reflection coefficient on the real axis,
\begin{equation*}
	x_j(\xi)=r\left(0,\xi;\{a_j^k\}_{k=1}^K\right),
\end{equation*}
on a uniform spectral grid $\xi\in[-5,5]$ with $M_\xi=4096$ points. Unlike the discrete Fourier transform, the IST continuous spectrum is not tied to a Nyquist interval set by the sampling rate in $t$: the reflection coefficient $r(\xi)$ is defined for all $\xi\in\mathbb{R}$, and numerical NFT implementations therefore choose a finite spectral window $[-\Xi,\Xi]$ and a grid resolution to balance truncation error and computational cost. In our experiments we set $\Xi=5$ because, for the considered RZ pulse trains, $|r(\xi)|$ is strongly concentrated near $\xi=0$ and becomes negligible outside this window; enlarging $\Xi$ did not change the reconstructed waveforms within the reported accuracy while increasing runtime and memory. The same rationale motivates the smaller window $[-1.5,1.5]$ in the end-to-end experiment, where we focus resolution on the spectral region carrying most of the energy for this signal class.  Numerically, $y_j(t)$ is sampled on a time grid of size $M_t$, while $r(\xi)$ is evaluated on a chosen spectral grid of size $M_\xi$ over a prescribed interval; these discretizations are selected independently to meet accuracy and cost requirements. A formal Nyquist-type analogue in this framework would require a sampling theory formulated for the relevant operator spectral representation, which is outside the scope of this work.

We then downsample both $x_j(\xi)$ and $y_j(t)$ to $n=m=1024$ points, forming the input-output pairs $(\mathbf{x}_j,\mathbf{y}_j)$ for the affine model dataset constructing them in matrices. In total, we generate $N=1000$ samples, covering approximately $1.5\%$ of the $2^{16}=65536$ possible bit patterns. These samples are split into training (70\%) and validation (30\%) sets.

The discretizations used here are intentionally oversampled relative to the practical symbol rate. On the time window $t\in[-90,90]$ and with bit-slot width $T=10$, the signal spans about $180/T\approx 18$ slots, so the downsampled representation with $m=1024$ corresponds to roughly $m/18\approx 57$ samples per slot. In practical coherent receivers, the digitization is typically on the order of a few samples per symbol (often $\sim 1$--$2$ samples/symbol), and therefore much smaller $m$ would be required for deployment.

We use larger $n$ and $m$ in this proof-of-concept study for two reasons. First, starting from a dense grid helps isolate the behavior of the affine surrogate from numerical artifacts of the scattering computation. Second, the objective here is to demonstrate the low-rank structure and the existence of an accurate reduced representation, rather than to optimize the minimal sampling budget. Studying how reconstruction accuracy degrades as $n$ and $m$ are reduced toward practical sampling rates, and developing discretization robust (mesh-independent) variants of the learned operator, requires additional study. A promising direction for such mesh-independent reduced operators is to exploit structured low-rank representations (like tensor-train decompositions) for the discretized map and related kernels, which can yield compressed representations with favorable scaling as the discretization size increases~\cite{oseledets2011tensor}.

Before finding the affine map, we normalize each dataset by subtracting the training‐set mean and dividing by its standard deviation, computed separately for the scattering data (inputs) and the initial potentials (outputs). The same training‐set scaling factors are then applied to the validation set. During the prediction phase, the outputs of the trained affine map are denormalized back to original scales before computing the loss.

\subsection{Numerical experiments}
\label{sec:num-exp}

In our numerical experiments, we observed that using only the real (or alternatively the imaginary) part of the reflection coefficient $x_j$ is sufficient to construct an accurate affine approximation model. All results reported below are based on the real part of $x_j$.

This is an empirical observation for the restricted signal family considered here. Within this family, the reflection coefficients occupy a low-dimensional subset of the complex function space, so either real or imaginary part of reflection coefficient alone remains sufficiently informative to recover $u(0,t)$ with small relative error. An intuitive explanation is that, for considered pulse trains, the scattering data exhibit strong consistency constraints, so the two real and imaginary parts of $r(\xi)$ are not statistically independent on this dataset.

First, we applied Eq.~(\ref{eq:full-map}) to the training data to compute the full-order operator $A_{\mathrm{aug}}$ and the bias vector $\mathbf{b}$. Then, by computing the singular value decomposition of $A_{\mathrm{aug}}$, we observed a sharp decay, over several orders of magnitude, in the singular values after the sixteenth singular value. This indicates that a rank-16 approximation captures all of the operator's significant components. Accordingly, we set $r=16$ when constructing the projection matrix $R$ and the reduced mapping matrix $A_r$. This choice is also consistent with a natural lower bound imposed by the signal model: our dataset consists of $K=16$ bit slots (pulses), with each slot carrying an independent symbol $a^k\in\{\pm1\}$. Therefore the family of transmitted waveforms is parameterized by $K=16$ independent degrees of freedom, and any reduced representation intended to preserve these degrees of freedom (i.e., avoid collapsing distinct bit patterns to the same reduced coordinates) must have dimension at least $K$.

In practice, the transmitted data stream is typically organized into finite length blocks (packets/frames) of a fixed size. The parameter $K$ in our signal model can therefore be interpreted as the block length processed jointly by the integrability-based receiver in a single instance of the inverse map. Increasing $K$ improves spectral efficiency, but it also changes the dimensionality and conditioning of the inverse map and may affect the singular value decay of the learned operator. A systematic study of how reconstruction accuracy and the effective rank scale with the block length $K$ (including the trade-off between throughput and numerical stability) requires additional study.

Using reduced order affine map~(\ref{eq:reduced-map}), we generated predictions on both the training and validation sets. We quantify model performance via the relative $L_2$ error defined in Eq.~(\ref{eq:loss}). On the training set (700 samples) and the validation set (300 samples), the average losses are
\begin{equation*}
	\langle C\rangle_{\mathrm{training set}}\approx1.9\times10^{-6}
	\quad\text{and}\quad
	\langle C\rangle_{\mathrm{validation set}}\approx7.5\times10^{-5}.
\end{equation*}

\begin{figure}
	\centering
	\includegraphics[width=\linewidth]{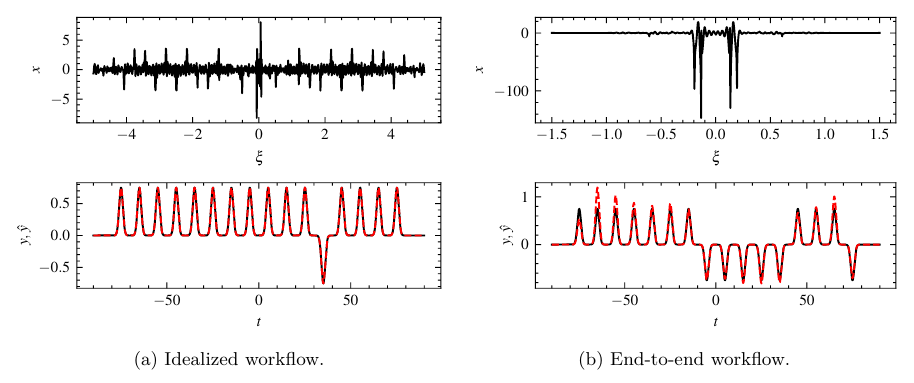}
	\caption{Worst case reconstructions for two settings. (a) Idealized workflow: the scattering data are computed directly from the transmitted waveform at $z=0$. Top: real part of the reflection coefficient $x=r(0,\xi)$; bottom: comparison of the ground truth initial potential $y$ (solid black line) and the affine reconstruction $\hat y$ (red dashed line), with $C\approx 1.3\times10^{-3}$. (b) End-to-end workflow: NLSE propagation to $z=z_{\mathrm{end}}$, computation of scattering data at $z=z_{\mathrm{end}}$, spectral backpropagation of the continuous spectrum to $z=0$, and affine reconstruction. Top: real part of the backpropagated reflection coefficient $x=r_{\mathrm{bp}}(0,\xi)$; bottom: comparison of $y$ (solid black line) and $\hat y$ (red dashed line), with $C_{\mathrm{worst}}\approx 2.2\times10^{-1}$. The reconstruction in (b) is visibly less accurate than in (a) due to compounded numerical errors from NLSE propagation, estimating scattering data from the propagated waveform, and the subsequent spectral backpropagation. Nevertheless, the reconstructed pulse phases (signs) and pulse locations are recovered correctly, which is sufficient for bit identification even when some pulse amplitudes are not reconstructed perfectly.}
	\label{fig:worst_cases}
\end{figure}

The relative $L_2$ loss in Eq.~(\ref{eq:loss}) is a normalized waveform discrepancy, thus $C\ll1$ means that the reconstruction error is small compared to the overall signal energy. In this regime, the recovered temporal profile preserves the amplitude and phase structure of the pulse train and, visually, the reconstructed waveform is essentially indistinguishable from the target at the scale of the signal (see Fig.~\ref{fig:worst_cases}a). 

To assess whether the small losses could be an artifact of overfitting, we report performance on an independent validation set that was not used to identify the operator parameters. The fact that the validation loss remains comparably small indicates that the learned affine operator generalizes to unseen bit patterns within the same signal class, rather than memorizing the training examples.

A fully application-level notion of sufficient accuracy is typically expressed in terms of the bit error rate (BER) after symbol decision. However, BER depends on the receiver decision rule and on the assumed noise and hardware model. A rigorous BER characterization therefore requires an end-to-end communication setup and is beyond the present proof-of-concept study. An end-to-end variant of this experiment, in which the affine map is embedded into an NLSE propagation and scattering data backpropagation workflow, is presented in Sec.~\ref{subsec:end_to_end}.

Figure~\ref{fig:worst_cases}a illustrates the worst case prediction on validation set, which corresponds to the highest error ($C \approx 1.3\times10^{-3}$) over this set, along with the associated reflection coefficient.

\begin{figure}
	\centering
	\begin{subfigure}{0.49\linewidth}
		\centering
		\includegraphics[width=\textwidth]{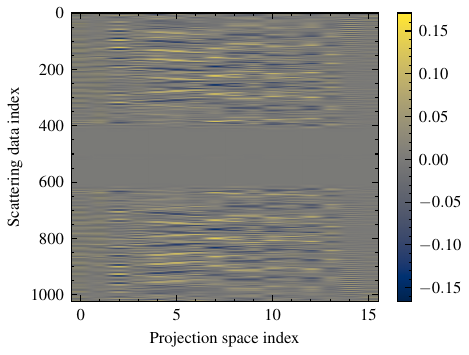}
		\caption{Projection into an $r$-dimensional subspace via $R$}
		\label{fig:projection_matrix}
	\end{subfigure}
	\begin{subfigure}{0.49\linewidth}
		\centering
		\includegraphics[width=\textwidth]{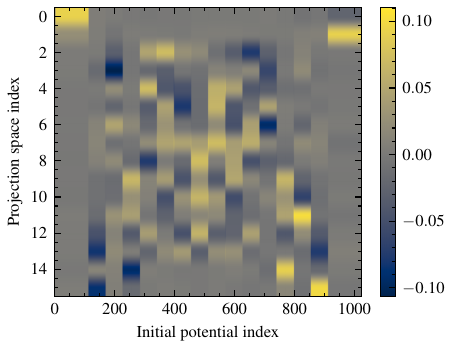}
		\caption{Low-rank mapping $A_r$ from reduced input to output}
		\label{fig:mapping_matrix}
	\end{subfigure}
	\caption{Low-rank decomposition of the affine map: (a) projection matrix and (b) reduced mapping matrix.}
\end{figure}

Figures~\ref{fig:projection_matrix}
and~\ref{fig:mapping_matrix}
illustrate the structure of the learned projection matrix $R$ and the reduced mapping matrix $A_r$, respectively. These visualizations can be understood directly from the truncated-SVD construction $A_{\mathrm{aug}}\approx U_r\Sigma_r V_r^T$ used in Eq.~(\ref{eq:reduced-map}). In this factorization, the matrix $R=U_r\Sigma_r$ (Fig.~\ref{fig:projection_matrix}) plays the role of an analysis operator: it maps the sampled reflection coefficient  into a vector of reduced coordinates $\mathbf{c}=\mathbf{x}_{\mathrm{aug}}R\in\mathbb{R}^r$. Each column of $R$ therefore defines a weighted linear functional on the scattering grid, i.e., a feature extractor that combines information from many $\xi$-samples. The visible banded (row-wise) structure in Fig.~\ref{fig:projection_matrix} indicates that the dominant reduced coordinates are not arbitrary mixtures: each mode emphasizes specific spectral regions, reflecting structured correlations in $r(\xi)$ produced by the underlying pulse-train family.

Conversely, the reduced mapping $A_r=V_r^T$ (Fig.~\ref{fig:mapping_matrix}) acts as a synthesis operator that turns the reduced coordinates into the time-domain signal $\hat{\mathbf{y}}=\mathbf{c}A_r$. The prominent feature in Fig.~\ref{fig:mapping_matrix} is the appearance of approximately $K=16$ vertical column strips. The target waveform is a sum of $K=16$ well-separated Gaussian pulses located in distinct bit slots. Since the output index is ordered in time, samples belonging to the same slot form blocks; the map therefore organizes itself, producing strip-like regions that correspond one-to-one with the $16$ pulses. In other words, $A_r$ is close to a block structured (slot local) map: the reconstruction is assembled from contributions that are localized around each pulse position, with signs encoding the $\pm1$ DPSK symbols. Thus, the decomposition $A_{\mathrm{aug}}\approx R A_r$ yields a two-stage surrogate for the inverse scattering step: project the continuous spectrum onto a small set of dominant modes, and synthesize the time-domain pulse train from these modes.

\subsection{End-to-end workflow simulation}
\label{subsec:end_to_end}

The primary objective of this work is to construct an affine map surrogate for the inverse scattering solution from scattering data to the corresponding initial potential. In \cref{sec:data-gen,sec:num-exp} we evaluate this surrogate in an idealized setting, where the scattering data are computed directly from the transmitted waveform at $z=0$. Here we embed the same surrogate into an end-to-end, integrability-based signal processing workflow: the waveform is propagated by the NLSE to $z=z_{\mathrm{end}}$, the scattering data are computed at $z=z_{\mathrm{end}}$, the continuous spectrum is backpropagated to $z=0$, and the affine surrogate is found and applied to recover an approximation of the transmitted signal. The full procedure is summarized in Algorithm~\ref{alg:end_to_end}.

Compared to the idealized inverse scattering map discussed in \cref{sec:data-gen,sec:num-exp}, this end-to-end experiment incorporates additional numerical effects due to NLSE propagation, estimation of scattering data at the propagation end, and spectral backpropagation. A systematic treatment and reduction of these additional error sources is beyond the scope of this work. Nevertheless, the results below show that even this implementation yields accurate reconstructions, indicating that the affine map approach is robust to moderate imperfections.

\begin{algorithm}[t]
	\caption{End-to-end NLSE propagation, spectral backpropagation, and affine map reconstruction}
	\label{alg:end_to_end}
	\begin{algorithmic}[1]
		\For{each realization}
		\State Generate a transmit waveform $u(0,t)$ as an RZ Gaussian pulse train.
		\State Propagate $u(0,t)$ to $u(z_{\mathrm{end}},t)$ by direct numerical simulation of the NLSE.
		\State Compute the reflection coefficient $r(z_{\mathrm{end}},\xi)$ for $u(z_{\mathrm{end}},t)$.
		\State Backpropagate the continuous spectrum to $z=0$ to obtain $r_{\mathrm{bp}}(0,\xi)$.
		\State Reconstruct an approximation $\hat u(0,t)=G_\theta \bigl(r_{\mathrm{bp}}(0,\xi)\bigr)$ using the affine map.
		\EndFor
	\end{algorithmic}
\end{algorithm}

To reduce finite window artifacts during propagation, including pulse broadening due to dispersion, we simulate the NLSE on an enlarged time window. Specifically, the input pulse train is generated on a small grid $t\in[-90,90]$ with $M_t=4096$ points (as in \cref{sec:data-gen}), embedded into a big grid with the same $\Delta t$ via temporal padding (here, a padding factor of $4$), and propagated on the big grid $t\in[-360,360]$. Propagation is performed with an integrating-factor method~\cite{yang2010nonlinear} for the focusing NLSE Eq.~\eqref{eq:nlse} using step size $\Delta z=0.1$ up to $z_{\mathrm{end}}=50$ (normalized units). We monitor boundaries by tracking the fraction of signal energy contained near the window edges after propagation; the chosen padding keeps this quantity negligible for the realizations considered.

\begin{figure}
	\centering
	\includegraphics[width=1.0\linewidth]{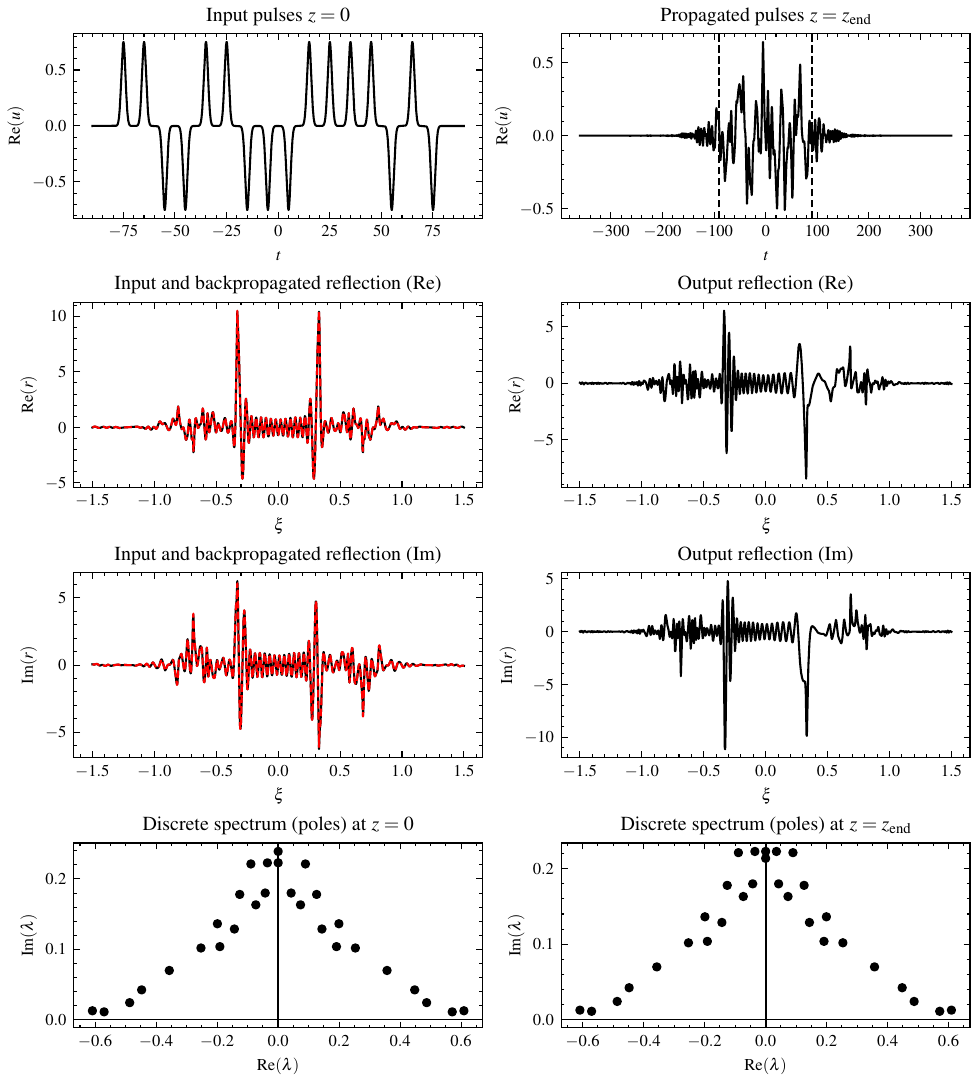}
	\caption{End-to-end representative realization. The plot shows the input waveform on the small window ($z=0$), the propagated waveform on the padded window ($z=z_{\mathrm{end}}$) with the crop interval indicated, the continuous spectrum (reflection coefficient) at $z=0$ and $z=z_{\mathrm{end}}$ together with its backpropagated version, and the discrete eigenvalues computed at $z=0$ and at $z=z_{\mathrm{end}}$.}
	\label{fig:end-to-end}
\end{figure}

For each realization we compute the scattering data using the same FNFT implementation as in \cref{sec:data-gen}. The continuous spectrum is represented by the reflection coefficient $r(z,\xi)$ sampled on a uniform grid $\xi\in[-1.5,1.5]$. We use a narrower interval in the end-to-end experiment to focus resolution on the spectral region carrying most of the energy for the considered signal class. In addition, we explicitly compute the discrete spectrum, i.e., eigenvalues $\{\lambda_k\}$ in the upper half-plane $\Im\lambda_k>0$ at which $a(z,\lambda_k)=0$. Figure~\ref{fig:end-to-end} shows a representative realization and includes: the input and propagated waveforms; the continuous spectrum at $z=0$ and $z=z_{\mathrm{end}}$; the discrete eigenvalues computed at both ends. The discrete spectral components are present for the considered RZ Gaussian pulse-train family, even though the reconstruction uses only the continuous spectrum.

We backpropagate the continuous spectrum computed at $z=z_{\mathrm{end}}$ to $z=0$ by undoing its phase evolution
\begin{equation*}
	r_{\mathrm{bp}}(0,\xi)
	=
	r(z_{\mathrm{end}},\xi) 
	\exp \Bigl(-i  4  d  z_{\mathrm{end}} \xi^2\Bigr),
\end{equation*}
where $d=0.5$ is the dispersion coefficient used in the propagation solver. The agreement between $r_{\mathrm{bp}}(0,\xi)$ and the reflection coefficient computed directly from $u(0,t)$, as shown in Fig.~\ref{fig:end-to-end}, provides a numerical consistency check for the DST and backpropagation part of workflow.

We train an affine surrogate to map the backpropagated continuous spectrum to the original input waveform,
\begin{equation*}
	\hat u(0,t)
	=
	G_\theta \bigl(r_{\mathrm{bp}}(0,\xi)\bigr),
\end{equation*}
using the same procedure as in \cref{sec:num-exp}. In this end-to-end setting the affine map operates on scattering data that have undergone propagation, DST, and backpropagation; consequently the reconstruction accuracy reflects the compound numerical error of the full workflow rather than the inverse scattering approximation alone. In this experiment, the training pairs are $\bigl(r_{\mathrm{bp}}^{(j)}(0,\xi), u^{(j)}(0,t)\bigr)$, so the affine map is identified directly for the receiver side processed input.

For $N=1000$ realizations (70\% training, 30\% validation), the relative $L_2$ losses are
\begin{equation*}
	\langle C\rangle_{\mathrm{training set}} \approx 3.1\times 10^{-13},
	\qquad
	\langle C\rangle_{\mathrm{validation set}} \approx 1.4\times 10^{-2},
\end{equation*}
and the worst validation example attains $C_{\mathrm{worst}}\approx 2.2\times 10^{-1}$. Figure~\ref{fig:worst_cases}b shows the worst case validation sample: the backpropagated continuous spectrum used as input and the corresponding time-domain waveform reconstruction. The remaining mismatch is mainly due to compounded numerical errors in the end-to-end simulation, including NLSE propagation on a finite window, numerical evaluation of the scattering data for the propagated waveform, and the subsequent spectral backpropagation. Despite these imperfections, the reconstruction preserves the pulse locations and the relative phase pattern (signs) across slots, which is sufficient for reliable bit identification even when some pulse amplitudes are not recovered exactly.

The discrete spectrum for RZ Gaussian pulse trains is not fixed: depending on the particular bit pattern and the overall amplitude scale, the Zakharov-Shabat problem may have zero, one, or multiple discrete eigenvalues. In the examples shown in Fig.~\ref{fig:end-to-end}, discrete spectrum is present and visualized at both $z=0$ and $z=z_{\mathrm{end}}$, indicating the coexistence of continuous and discrete spectral components for considered signal family and parameter regime. In this workflow the reconstruction uses only the continuous spectrum.

\subsection{Computational cost}
The DST is first evaluated on a dense grid ($M_t=M_\xi=4096$) to ensure numerical accuracy of the scattering data and to avoid resolution artifacts. The affine surrogate, however, is intended to act on discretized receiver data with finite sampling rates. For this reason we downsample both the reflection coefficient and the time-domain signal to $n=m=1024$, which is closer to practical discretizations while still remaining deliberately relatively large for this proof-of-concept study.

It is important to distinguish between the training stage and the inference stage. Let $X_{\mathrm{aug}}\in\mathbb{R}^{N_s\times d}$ denote the augmented input matrix, with $N_s$ the number of training examples and $d=2n+1$. Computing the Moore--Penrose inverse (using an SVD-based routine) scales as $O(N_s d^2)$ when $N_s\ge d$ and as $O(d N_s^2)$ when $d\ge N_s$, i.e., cubic in the smaller dimension. This cost is paid once during training. In contrast, the online evaluation of the dense affine map $\hat y = x_{\mathrm{aug}}A_{\mathrm{aug}}$ with $A_{\mathrm{aug}}\in\mathbb{R}^{d\times 2m}$ is a matrix--vector multiplication with cost $O(dm)$, which becomes $O(N^2)$ when $n\sim m\sim N$. Using the reduced-rank factorization $A_{\mathrm{aug}}\approx R A_r$ with $r\ll n,m$ lowers the online cost to $O(dr)+O(rm)=O(r(n+m))\sim O(rN)$, which is the main mechanism that keeps inference efficient as the discretization grows.

Finally, fast nonlinear Fourier transform (FNFT) provides an efficient DST computation with complexity $O(N\log^2N)$ in the number of samples $N$. Our surrogate targets the inverse map from (backpropagated) scattering data to the time-domain waveform and is designed to replace repeated inverse scattering solvers with a simple linear operator application at the receiver.

\section{Discussion}\label{sec:discussion}
We note that our workflow intentionally excludes the discrete spectrum component of the scattering data. At the same time, we verified numerically that discrete eigenvalues are indeed present for a subset of the generated waveforms (see Fig.~\ref{fig:end-to-end}). The present contribution is that, for this signal class and parameter regime, the continuous spectrum alone is sufficient for recovery with high fidelity. This simplification enables efficient approximations of both the direct and inverse scattering transforms for this class of signals, which is particularly advantageous in high-speed optical communication systems. Intuitively, the reflection coefficient implicitly encodes information about the discrete part, allowing us to skip an explicit use of the discrete spectrum.

Furthermore, our numerical experiments demonstrate that, for Gaussian pulse trains, using only the real or the imaginary part of the reflection coefficient still yields an accurate affine map approximation. This observation further reduces computational complexity and data requirements.

The present study does not explicitly model additive noise sources such as amplified spontaneous emission (ASE) from optical amplifiers, nor structural fiber disorder; these effects will be addressed in future studies. Instead, we test the surrogate under a different but practically relevant mechanism that produces strong, effectively random distortions within the pulse train: nonlinear interactions between localized pulse components and excess continuous radiation. Such interactions are known to induce phase distortions, energy redistribution along the sequence (amplitude jitter), and pulse position fluctuations (time jitter) because the pulses are not fully transparent to the accompanying radiation field~\cite{kuznetsov1995nonlinear, chung2004interaction, chertkov2003shedding}. In this sense, the perturbation source is the pseudo-random structure of the transmitted sequence itself, and the end-to-end experiment in Sec.~\ref{subsec:end_to_end} demonstrates that the affine map surrogate remains effective under these nonlinear interaction effects. A systematic assessment under ASE and structural disorder, including BER-level impact and training that accounts for noise, requires additional study.

The affine map approximation is especially well suited for implementation on dedicated hardware since it relies only on simple linear algebra operations. In addition, the reduced order affine model suggests a pathway for reduced order model of DST algorithms, such as those based on the Boffetta-Osborne method~\cite{boffetta1992computation}, the fast nonlinear Fourier transform~\cite{wahls2013introducing}, or numerical Riemann-Hilbert solvers~\cite{trogdon2013numerical}.

A practical challenge arises from the behavior of the reflection coefficient~(\ref{eq:ref-coef}), which exhibits sharp peaks along the real axis when a discrete spectral pole is located near ($\Im \lambda$ close to 0 for discrete pole at $\lambda$). Although $a(z,\xi)\neq 0$ for all real $\xi$, these near-pole effects cause large magnitudes and steep local gradients in $r(z,\xi)$. Consequently, the sampled values of the reflection coefficient can vary significantly depending on whether a grid point coincides with a peak or is slightly offset. This grid sensitivity should be addressed to ensure robust numerical performance.

\section{Conclusion}

We have presented an affine map operator as a computationally efficient surrogate for the inverse scattering transform, and demonstrated that its low-rank structure admits a compact reduced order model. Our numerical experiments show that, for return-to-zero Gaussian pulse trains, retaining only the continuous part of the Lax spectrum (reflection coefficient) suffices to reconstruct the original signal with high fidelity. This finding supports a fast, integrability-based signal processing framework with clear potential for real-time implementation in high-speed optical communication systems. Future work will address the development of mesh-independent operators to mitigate grid sensitivity, as well as extend framework to other modulation formats like non-return-to-zero.

\bibliographystyle{unsrtnat}
\bibliography{references}

\end{document}